\documentstyle[12pt,epsf,epsfig]{article}
\newcommand{\be} {\begin{equation}}
\newcommand{\ee} {\end{equation}}\newcommand{\ba} {\begin{eqnarray}}
\newcommand{\ea} {\end{eqnarray}}

\begin{document}
\newcommand{\ket}[1] {\mbox{$ \vert #1 \rangle $}}
\newcommand{\bra}[1] {\mbox{$ \langle #1 \vert $}}
\newcommand{\bkn}[1] {\mbox{$ < #1 > $}}
\newcommand{\bk}[1] {\mbox{$ \langle #1 \rangle $}}
\newcommand{\scal}[2]{\mbox{$ \langle #1 \vert #2  \rangle $}}
\newcommand{\expect}[3] {\mbox{$ \bra{#1} #2 \ket{#3} $}}
\newcommand{\ki}{\mbox{$ \ket{\psi_i} $}}
\newcommand{\bi}{\mbox{$ \bra{\psi_i} $}}
\newcommand{\p} \prime
\newcommand{\e} \epsilon
\newcommand{\la} \lambda
\newcommand{\om} \omega   \newcommand{\Om} \Omega
\newcommand{\cc}{\mbox{$\cal C $}}
\newcommand{\w} {\hbox{ weak }}   
\newcommand{\al} \alpha
\newcommand{\bt} \beta 

\def\lrD{\mathrel{{\cal D}\kern-1.em\raise1.75ex\hbox{$\leftrightarrow$}}}

\def\lr #1{\mathrel{#1\kern-1.25em\raise1.75ex\hbox{$\leftrightarrow$}}}

\overfullrule=0pt \def\sqr#1#2{{\vcenter{\vbox{\hrule height.#2pt
          \hbox{\vrule width.#2pt height#1pt \kern#1pt
           \vrule width.#2pt}
           \hrule height.#2pt}}}}
\def\square{\mathchoice\sqr68\sqr68\sqr{4.2}6\sqr{3}6}
\def\lrpartial{\mathrel
{\partial\kern-.75em\raise1.75ex\hbox{$\leftrightarrow$}}}

\overfullrule=0pt \def\sqr#1#2{{\vcenter{\vbox{\hrule height.#2pt
          \hbox{\vrule width.#2pt height#1pt \kern#1pt
           \vrule width.#2pt}
           \hrule height.#2pt}}}}
\def\square{\mathchoice\sqr68\sqr68\sqr{4.2}6\sqr{3}6}  
\def\lrpartial{\mathrel
{\partial\kern-.75em\raise1.75ex\hbox{$\leftrightarrow$}}}

\def\lrD{\mathrel{{\cal D}\kern-1.em\raise1.75ex\hbox{$\leftrightarrow$}}}

\def\lr #1{\mathrel{#1\kern-1.25em\raise1.75ex\hbox{$\leftrightarrow$}}}
\begin{flushright}
October 10-th, 1997
\end{flushright}
\vskip .5 truecm
\centerline{\Large\bf{The validity of the }}
\vskip 2 truemm
\centerline{\Large\bf{Background Field Approximation}}
\vskip 1. truecm

\centerline{{\bf R. Parentani}}
\vskip 3 truemm
\centerline{
Laboratoire de Math\'ematiques et Physique Th\'eorique, UPRES A 6083 CNRS}
\centerline{Facult\' e des Sciences, Universit\'e de Tours, 37200 Tours, France}
\centerline{
e-mail: parenta@celfi.phys.univ-tours.fr}

\vskip 1.5 truecm
\centerline{\bf Abstract }

In the absence of a tractable theory of quantum gravity, quantum matter field
effects have been so far
computed by treating gravity at the Background Field Approximation.
The principle aim of this paper is to investigate 
the validity of this approximation which is not specific to gravity.
To this end, for reasons of
 simplicity and clarity, we shall compare the descriptions of
thermal processes induced by constant acceleration (i.e. the 
Unruh effect)
in four dynamical frameworks. In this problem, the position of the ``heavy'' accelerated
system plays the role of gravity.
In the first framework, the trajectory is treated 
at the BFA: it is given from the outset and unaffected by radiative processes.
In the second one, recoil effects induced by these emission processes
are taken into account by describing the system's position by WKB wave functions.
In the third one, the accelerated system is described by 
second quantized fields and
in the fourth one, gravity is turned on. 
It is most interesting to see when and why 
transitions amplitudes evaluated in different
frameworks but describing the same process do agree.
It is indeed this comparison that determines the validity of the BFA.
It is also  interesting to notice that the abandonment of the BFA
delivers new physical insights concerning the processes. 
For instance, in the fourth framework,
the ``recoils'' of gravity show that
the acceleration horizon area acts as an entropy in delivering heat to 
accelerated systems.

\newpage

\section{Introduction}

In order to analyze the stability of a given configuration, one must choose
a framework in which the dynamical consequences
of certain class of fluctuations may be investigated. It is quite obvious that the
stability conditions might drastically depend on the class of fluctuations considered.
Moreover, in view of the complexity of most physically interesting situations, 
certain simplifying approximations must be made. These in turn specify 
the particular dynamical framework 
and may therefore restrict the class of fluctuations one is
effectively taken into account. Because of this, stability conditions 
might also depend on the nature of the approximations used.
In particular, upon dealing with gravitational systems, the absence 
of any tractable quantum 
theory of gravity imposes to treat gravity at the 
Background Field Approximation (BFA)
even when one considers quantum effects induced by 
quantum matter fields.
This implies that one can only compute the gravitational 
response induced by the mean (i.e.
quantum average) energy-momentum tensor. 
One must thus inquire onto the validity of the 
results obtained 
by using the BFA for describing gravity. 
In particular, when one abandons the BFA, does one obtain
a behavior more or less singular ?

These considerations certainly apply to unstable 
or near unstable situations. Examples 
are provided by matter and gravity dynamics near inner horizons of 
charged black holes\cite{israel}\cite{ori} as well as the dynamics near the event horizon of 
extremal black holes\cite{t}\cite{teddy}.
However, they might also apply to (apparently) more stable situations, 
such as, for instance,
the evaporation process of Schwarzschild holes\cite{Hawk}. 
This is because the whole dynamical
evolution might drastically depend on the nature of the approximations.
%
Indeed, when working in the semi-classical approximation, one only deals
with regular mean energy densities\cite{Bard}\cite{PP}\cite{Massar2}. 
However, 
Hawking quanta issue, in usual quantum field theory,
from vacuum fluctuations characterized by ultra-high frequencies (trans-Planckian), 
see \cite{THooft}-\cite{bmps}.
These fluctuations give rise to trans-Planckian 
matrix elements of the
stress energy-tensor\cite{mapa2}\cite{Verl3}. 
Therefore, since these matrix elements should determine (yet unknown)
quantum gravitational back-reaction effects,
the evolution in quantum gravity might strongly differ 
from the semi-classical one.
 
In this article, we shall investigate the relationships
between the choice of the dynamical framework and the description
of processes. To this end, we shall compare the descriptions of 
thermal effects induced by uniform acceleration, 
i.e. the Unruh\cite{Unr} effect,
in four dynamical frameworks.
This comparative analysis clearly reveals the {\it generic} properties
of the use of the BFA.
(By generic we mean that these properties should also apply to 
quantum gravity.)
In particular, it shows precisely when and why
 the description of processes obtained by using the BFA coincides 
to the description of the same processes in new frameworks in which the
dynamics has been enlarged.

 In order to perform this analysis,
one should carefully choose
the dynamical objects that shall be compared.
We shall see that the comparison is the most instructive 
when focused on specific matrix
elements controlling transition amplitudes\cite{rec}. 
Indeed these are the basic
elements out of which all physical quantities can be computed. Furthermore
they are very sensitive to the approximations used. Instead, integrated or averaged
quantities, such as transition rates or the total energy emitted, are much less sensitive
and therefore not appropriate to perform the comparison.
Notice that the (relative) insensitivity of the transition rates is related to 
the thermodynamical nature of the Unruh effect. 

The dynamical arena of the four frameworks we shall consider
will be progressively enlarged and will incorporate 
the former one.
Correspondingly, the description of the transitions will become more
sophisticated.
However, 
the enlargement of the dynamics 
shall provide new physical insights. In particular, the energy conservation law
changes drastically when one abandons the BFA. In addition to this 
change which is not specific to the Unruh effect, there are also
specific new informations obtained by the enlargement of the dynamics. 
The most interesting ones concern the strong relations between the Unruh effect
and both pair creation in a constant electric field (in the third framework) 
and gravitational entropy (in the fourth one). 

The first framework corresponds to the Unruh's original 
formulation of the problem\cite{Unr}. 
In this case, one deals with a uniformly accelerated 
two-level atom coupled to a quantum massless field in Minkowski space time.
The trajectory of the atom is treated at the BFA, i.e. it is 
once for all given and insensitive to the radiative events.
Thus one can neither analyze the stability of the trajectory
nor answer the
question of the origin of the energy emitted by the accelerated 
system since 
momentum conservation is violated during 
the emission processes.

In the second framework, the position of the accelerated system is quantized.
The particular model we shall consider consists of a ``heavy'' two-level ion immersed
in a constant electric field and coupled, as before, to the massless field\cite{BPS}\cite{rec}. 
In this second framework,
the momentum and energy conservation laws are fully respected
and one verifies 
that it is the electric field that provides the energy of the radiated quanta. 
We shall then establish under which conditions and for which reason the Unruh thermalization 
process is recovered in this new framework. The conditions are that both
(i) the WKB approximation of the wave functions
describing the system's position and 
(ii) a first order expansion in the energy changes induced by the transition should
be valid. Then,
Hamilton-Jacobi  equations imply that the transition {\it rates} 
computed from both frameworks agree.
(This agreement is absolutely 
generic in character: it also applies
to quantum gravity when gravity plays the role of the heavy system\cite{wdw}.)
This does not mean however that 
the transition {\it amplitudes} themselves coincide: there are indeed trans-Planckian
phase shifts induced by exponentially large Doppler shifts.
As a consequence, the mean flux radiated the accelerated system 
is smooth out by these recoil effects\cite{rec}. 
One thus realizes
that the singular energy density obtained by treating 
the trajectory at the BFA\cite{Unruh92}\cite{mapa1}
 was an {\it artifact} directly induced by the use of 
this approximation which ignored recoils.


However this second framework is still an approximation since pair creation
effects have not been taken into account. Indeed, since one is working with 
quantized relativistic fields, it is mandatory
to work in second quantization.
In this third framework, 
the new enlargement of the dynamics makes therefore contact with
a an priori completely different phenomenon, namely pair creation
in a constant electric field, i.e. the Schwinger effect\cite{schw}.
It is remarkable that the amplitudes describing 
this latter effect are directly related to the radiative transitions 
by ``crossing symmetry''\cite{suh},
an intrinsic analytical property of amplitudes in QFT. 
Therefore, the Unruh effect and
the Schwinger effect should be conceived as two aspects of a single
dynamical theory, namely QFT in a constant electric field\cite{Niki, apple}.
In addition, this connection between pair creation amplitudes and 
radiative processes sheds light on the recently analyzed pair creation 
processes of charged black holes\cite{(11)}-\cite{(4)}. 
In particular, the role of the (gravitational) instanton
which relates by tunneling process the Melvin geometry (in which the black holes are absent)
to the Ernst geometry (present) is clarified\cite{gravitinst}.  

This last remark naturally introduces the dynamical role of gravity 
that has been ignored so far.
Indeed, upon turning on gravity, i.e. $G \neq 0$, the thermodynamical role of the 
acceleration horizon emerges.
Then, the Unruh effect is included
into the more general framework of thermodynamics of horizons. 
In particular, the role of the horizon area 
acting as an entropy in delivering heat to accelerated systems (which are 
at rest with respect to the
Killing horizon)
 is established. 
These relations between Euclidean gravity
and thermal phenomena can also be conceived as particular examples 
of the thermodynamical approach
to gravity\cite{jac} presented by Ted Jacobson.

In resume, by abandoning the BFA for describing the system's 
trajectory, we shall successively

\noindent
1.{ restore energy-momentum conservation,}

\noindent
2.{ suppress singular behavior of the emitted fluxes,}

\noindent
3.{ relate radiative transitions to pair creation amplitudes and}

\noindent
4.{ connect thermal effects induced by constant acceleration to
horizon entropy.}

Throughout the text, we shall also mention what are the 
properties revealed by the analysis of Unruh's effect that are
generic in character.

\section{The transition amplitudes in the absence of recoils}

In the original Unruh framework\cite{Unr}, 
the two level atom is maintained, for all times, on a single uniformly accelerated trajectory
\begin{equation} t_a (\tau) =a^{-1}
\mbox{sinh} a \tau \ ,\  z_a (\tau) =a^{-1} \mbox{cosh} a \tau
\label{acctraj}
\end{equation}
$a$ is the acceleration and $\tau$ is the proper time of the accelerated 
atom. We work for simplicity in Minkowski 
space time in $1+1$ dimensions.
The two levels of the atom are designated by $\ket{-}$ and $\ket{+}$ for the
ground state and the excited state respectively. The transitions from one state
to another are induced by the operators $A, A^{\dagger}$
\begin{eqnarray}
A \ket{-} = 0, \quad A \ket{+} = \ket{-} \nonumber\\
A^{\dagger} \ket{-} = \ket{+}, \quad A^{\dagger} \ket{+} = 0
\label{operA}
\end{eqnarray}

The atom is coupled to a massless scalar field $\phi(t,z)$. The 
Klein-Gordon equation is $(\partial^2_t - \partial^2_z ) \phi =0$ 
 and the general solution is thus \begin{equation}
\phi(U,V) = \phi(U) + \phi(V) \label{phiuv}
\end{equation}
where $U,V$ are the light like coordinates given by $U=t-z, V=t+z$.
The right moving part may be decomposed into plane waves:
\begin{equation} \varphi_{\om} (U) = {e^{-i\om U} \over \sqrt{4 \pi \om}}
\label{modeU}
\end{equation}
where $\om$ is the Minkowski energy. Using this basis,
the Heisenberg field 
operator $\phi$ reads
\begin{equation}
\phi(U)= \int_0^{\infty} d\omega  \left( a_{\om}
\varphi_{\om} (U) + a_{\om}^\dagger \varphi_{\om}^* (U)\right)
\label{phiU}
\end{equation}
where $a_{\om}, a_{\om}^\dagger$ are operators of destruction and 
creation of a Minkowski quantum of energy $\om$.
The coupling between the atom and the field is taken to be, see \cite{rec},  
\ba
H_{\rm int} (\tau)
&=&  g a 
\left( A e^{-i \Delta m\tau} + A^{\dagger}
e^{i \Delta m\tau} \right)\phi(U_a(\tau)) 
 \label{Hint}
\nonumber\\
&=&
g a \int_0^{\infty} \! d\om
\left[ \left( 
A e^{-i \Delta m\tau} + A^{\dagger}
e^{i \Delta m\tau} \right) \right. \nonumber \\
 &&\quad \quad \times  \left. 
\left( {a_{\om}{ e^{i{\om e^{-a\tau}}/a}}
\over \sqrt{4 \pi \om}} + {a_{\om}^\dagger {e^{-i\om e^{-a\tau}/a}}
\over \sqrt{4 \pi \om}} 
\right)\right]
\ea
where $U_a(\tau)=t_a(\tau) - z_a(\tau)= -e^{-a\tau}/a$.
Because of the locality of the coupling, only $\phi(U_a(\tau))$ evaluated along
the classical trajectory enters into the interaction.

The classically forbidden transition amplitude (spontaneous excitation) from
the ground state $\ket{-}\ket{0}$, where $\ket{0}$ is Minkowski vacuum,
to the excited state $\ket{+} \ket{1_{\om}}$ containing one
quantum of energy $\om$
(where $\ket{1_{\om}}= a^\dagger_{\om }\ket{0}$)
is
\begin{equation} 
B(\om,\Delta m, a) = \bra{1_{\om}}\bra{+}  e^{-i\int d\tau H_{\rm int}}
\ket{-}\ket{0}
\label{B}
\end{equation}
To first order in $g$, one finds
\begin{eqnarray}
B(\om,\Delta m, a) &=& -ig a \int_{-\infty}^{+\infty} d\tau \
e^{i \Delta m \tau}\ {e^{-i\om e^{-a\tau}/a} \over \sqrt{4 \pi \om}}
\nonumber\\
&=& ig \
\Gamma(-i {\Delta m / a}) \
 {(\om/a)^{i{\Delta m / a}} \over \sqrt{4 \pi \om}} \
e^{-\pi {\Delta m / 2a}}
\label{B1g}
\end{eqnarray}
where $\Gamma(x)$ is the Euler function.
This transition amplitude is closely related to the $\beta$ coefficient
of the Bogoliubov transformation\cite{Full}
 which relates the Minkowski operators
$a_{\om}$ to the Rindler operators associated with the eigenmodes of
$-i\partial_{\tau}$ (hence given by $\varphi_{Rindler}(\tau) = e^{-i\lambda
\tau}$). This provides the dynamical justification of studying Bogoliubov
coefficients which can be defined in the absence of any
coupling to accelerated system. Their study is thus preparatory in character,
exactly like the analysis of Green functions in free field theory. This point 
of view has been developed in \cite{wdw} to incorporate gravitational ``recoil'' effects in 
quantum cosmology along the same lines that recoil effects of accelerated systems
shall be treated in the next Section.

The transition amplitude for the inverse process,
i.e. disintegration from $\ket{+}\ket{0}$ to $\ket{-}\ket{1_{\om}}$, is given by
$A(\om,\Delta m, a) = B(\om, -\Delta m, a)$. From 
eq. (\ref{B1g}), one easily finds
\begin{equation} 
A(\om,\Delta m, a) = - B^*(\om,\Delta m, a ) \ e^{\pi \Delta m /a}
\label{ratio}
\end{equation}
Thus for all $\om$ one has
\begin{equation} 
\vert {B(\om,\Delta m, a) \over A(\om,\Delta m, a) }\vert^2 = e^{-2\pi \Delta m /a}
\label{ratio2}
\end{equation}
Since this ratio is independent of $\om$, the ratio of the 
probabilities of transitions (excitation and disintegration) is also given by eq. (\ref{ratio2}).
Hence at equilibrium, the ratio of the probabilities $P_-, P_+$
 to find the atom in the ground or excited state satisfy
\begin{equation} 
{P_+ \over P_-}= \vert {B(\om,\Delta m, a) \over A(\om,\Delta m, a)} \vert^2 = 
e^{-2\pi \Delta m /a}
\label{ratio3}
\end{equation}
This is the Unruh effect\cite{Unr}:
 at equilibrium, the probabilities of occupation
are thermally distributed with temperature $a/2 \pi$.

Using the amplitudes $B(\om,\Delta m, a)$ and $A(\om,\Delta m, a)$,
one can compute, to order $g^2$,  the mean value of the flux
emitted by the two level atom, see \cite{UnrW}-\cite{MPB}. 
The point we wish to emphasize is 
that all the emitted Minkowski quanta, whatever is their energy $\om$,
interfere so as to
deliver a negative mean flux, for $U < 0$, whose interpretation is that some Rindler
quantum has been absorbed\cite{grove}. However, the energy density 
is positive and singular on the
horizon $U=0$\cite{Unruh92}. This singular and physically pathological 
behavior arises from the 
extremely well tuned phases of $B(\om,\Delta m, a)$ and $A(\om,\Delta m, a)$ 
for $\om \to \infty$.
These phases
will be inevitably washed out after a finite proper time when recoils will be taken 
into account\cite{rec}. 
Then, the new value of the flux 
will be will be rapidly positive
and perfectly regular around and on $U=0$.

\section{The amplitudes in first quantization}

In this section, we first introduce the model of 
\cite{BPS} which is similar to the one used by Bell and 
Leinaas\cite{BL}. This will allow us
to take into account the momentum transfers to the accelerated system 
which are caused by the radiative emission processes.

We shall then establish when and why the thermalization of the inner degrees
of freedom of the accelerated system is recovered.
In particular, we shall see that the thermalization does not 
require a well defined
classical trajectory; therefore it neither requires well defined ``Rindler'' 
energies nor a well defined location of the horizon. Indeed, even when 
one deals with delocalized waves,
the 
thermal equilibrium ratio eq. (\ref{ratio3}) is obtained.

The model consist on two scalar charged fields ($\psi_M$ and $\psi_m$)
of slightly different
masses ($M$ and $m$)
 which will play the role of the former states of the atom: $
\ket{+}$ and $ \ket{-}$. The quanta of these fields are accelerated by an
external classical constant electric field $E$ with $a$ given by
\begin{equation}
{E \over M} = a \simeq {E \over m}
\label{accE}
\end{equation}
because one imposes
\begin{equation}
\Delta m = M - m <\!\!< M
\label{diffm}
\end{equation}
to have the ``light'' mass gap $\Delta m$ well separated from the ``heavy''
rest mass of the ion.

We work in the homogeneous gauge ($A_t=0$, $A_z= -Et$). In that gauge,
 the momentum $k$ is a conserved
quantity and the energy $p$ of a relativistic particle of mass $M$ is given by
the mass shell constraint $(p_{\mu}- A_{\mu})^2=M^2$:
\begin{equation}
p^2(M, k, t) = M^2 + (k + Et)^2
\label{KGE}
\end{equation}
The classical equations of motion are easily obtained from this equation
and are given in terms of the proper time $\tau$ by
\begin{eqnarray}
p(M,k, t) &=& M \mbox{cosh} a \tau \nonumber\\
t + k/E &=& (1/ a)  \mbox{sinh} a \tau \nonumber\\
z - z_0  &=& (1/ a) \mbox{cosh} a \tau
\label{eqmot}
\end{eqnarray}
Thus at fixed $k$, the time of the turning point (i.e. $dt/d\tau = 1$)
is fixed whereas its position is arbitrary and given by $z_0 +1/ a$.

From eq. (\ref{KGE}), 
the Klein Gordon equation for a mode $\psi_{k,M}(t,z)=
e^{ikz} \chi_{k,M}(t)$ is
\begin{equation}
\left[ \partial_t^2 + M^2 + (k+Et)^2 \right] \chi_{k,M}(t) = 0
\label{KGE2}
\end{equation}
When $\Delta m \simeq a $ and when eq. (\ref{diffm}) is satisfied, 
one has $M^2/E>\!\!>1$. Then pair production amplitudes\cite{schw}
may be safely ignored since the mean density of produced pairs
scales like $e^{-\pi M^2/E}$, see next Section. Furthermore, 
the WKB approximation for the modes  $\chi_{k,M}(t)$ is 
valid for all $t$. Indeed, the corrections to this approximation
are smaller than $(M^2/E)^{-1}$. 
The modes 
$\psi_{k,M}(t,z)$ can be thus correctly approximated by
\begin{equation}
\psi^{WKB}_{k,M}(t,z) = { e^{ikz}\over \sqrt{2 \pi}}{
e^{-i \int^t p(M, k, t^\prime ) dt^\prime} \over \sqrt{p(M, k, t)}} 
\label{WKB}
\end{equation}
where $p(M, k, t^\prime )$ is the classical energy at fixed $k$
given in eq. (\ref{KGE}).

As emphasized in refs. \cite{BPS}\cite{BMPPS}, the gaussian wave packets in $k$
do not spread if their width is of the order of $ E^{1/2}$. 
In that case, the spread  in $z$ at fixed $t$
is of the order of ${E}^{-1/2}=(M a)^{-1/2}$ and thus much 
smaller than the acceleration length $1/a$.
Since the stationary phase condition of the $\psi^{WKB}_{k,M}(t,z)$
modes gives back the accelerated trajectory  
and since
the wave packets do not spread, one has constructed a 
quantized version of the accelerated system which tends uniformly
to the BFA model when $M \to \infty, E \to \infty$ with $E/M= a$ fixed.

The interacting Hamiltonian which induces transitions between the quanta of
mass $M$ and $m$ by the emission or absorption of a massless neutral
quantum of the $\phi$ field is simply 
\begin{equation}
H_{\psi \phi} = \tilde g M^2 
\int dz \left[ \psi_M^\dagger (t,z)\psi_m (t,z) + 
\psi_M(t,z) \psi_m^\dagger (t,z) 
\right] \phi(t,z)
\label{inter2}
\end{equation}
where $\tilde g$ is dimensionless. 
In momentum representation, by limiting ourselves to the right moving 
modes of the $\phi$ field, one obtains
\begin{eqnarray}
H_{\psi \phi} &=& \tilde g M^2 \int_{-\infty}^{+\infty} dk \int_0^{\infty} 
{ d\om \over \sqrt{4 \pi \om}}
\nonumber \\
&& \quad \times
\left\{
\left[ b_{M,k-\om} \chi_{M,k-\om}(t) \ b^\dagger_{m,k}\chi^*_{m,k}(t) 
+ \mbox{h.c.} \right] a_{\om} {e^{-i\om t} } 
\right.
\nonumber\\
&&\quad \quad \left. +
\left[ b_{M,k+\om}\chi_{M,k+\om}(t) \ b^\dagger_{m,k}\chi^*_{m,k}(t)
+ \mbox{h.c.} \right] a_{\om}^\dagger {e^{+i\om t} }
\right\}\quad  \quad \quad 
\label{inter22}
\end{eqnarray}
where the operator $b_{M,k-\om}$ detroys
a quantum of mass $M$ and momentum
$k-\om$ and the operator $b^\dagger_{m,k}$ creates
a quantum of mass $m$ and momentum 
$k$. Therefore the product $b_{M,k-\om}b^\dagger_{m,k}$ plays
the role of the operator $A$ in eq. (\ref{Hint}).
However, the interaction between the 
radiation field $\phi$ and the two levels of the accelerated system is
no longer a priori restricted to a classical trajectory.
In the present context of homogeneous electric field, this leads to 
an exact conservation of momentum.
Notice that we have not introduced anti-ion creation operators in 
 $H_{\psi \phi}$. This is a legitimate
truncation when working with WKB waves. This condition shall be relaxed in 
the next Section.

We first establish when and why the behavior of the $\chi_{M}(t)$ modes re-delivers 
the Unruh effect\cite{rec}.
To this end, we shall not consider well localized wave packets. 
We emphasize this point. An alternative route, a priori
equally valid, would consist in {\it first} making well localized wave packets and
only {\it then} computing transition amplitudes, see \cite{BPS}. 
It turns out that the averaging over $k$
in constructing wave packets is both unnecessary and a nuisance in that it 
erases the detailed mechanisms that ensure thermalization in this new 
framework.

Thus, we focus on the transition amplitude at fixed momentum $k$
that replaces $B(\Delta m, \om, a)$, eq. (\ref{B1g}). 
In the present context, it corresponds to the amplitude to jump from the state
$\ket{1_k, m} \ket{0, M} \ket{0}$ to the state
$\ket{0,m} \ket{1_{k^\prime}, M} \ket{1_{\om}}$. $\ket{0, M}$ designates 
the vacuum state for the $\psi_M$ field and  $\ket{1_k, m} = b_{m, k}^\dagger
\ket{0,m }$ is the one particle state of the $\psi_m$ field of
momentum $k$.
Due to momentum conservation,
this amplitude can be expressed as
\begin{eqnarray}
 \delta (k &-& k^\prime - \om) \
\tilde B(M, m, k, \om) \nonumber \\
&=&  \bra{1_{\om}} \bra{1_{k^\prime},M} \bra{0,m}
e^{-i \int dt H_{\psi \phi}} \ket{1_k,m} \ket{0,M} \ket{0}
\label{calB}
\end{eqnarray}
compare this expression with eq. (\ref{B}).
\newline
To first order in $\tilde g$, 
one finds
\begin{eqnarray}
\tilde B(M, m, k, \om) = -i  \tilde g M^2 
 \int^{+\infty}_{-\infty}  dt \ \chi^*_{M, k - \om}(t) \ \chi_{m, k}(t) 
{e^{i \om t} \over \sqrt{4 \pi \om}}
\nonumber\\
=  -i \tilde g M^2 
 \int^{+\infty}_{-\infty} dt  {e^{i\int^t dt^\prime 
 \left[ p(M,k - \om, t^\prime) - p(m, k, t^\prime)
\right]} \over \sqrt{p(M,k - \om, t) p(m, k, t)}}
{e^{i\om t} \over \sqrt{4 \pi \om}}
\label{tildB2}
\end{eqnarray}
In the second line, we have used the WKB approximation eq. (\ref{WKB})
for the
$\chi$ modes.

As such, $\tilde B(M, m, k, \om)$ seems very different 
from 
the BFA amplitude $B(\Delta m, \om, a)$ given in eq. (\ref{B1g}).
Indeed, this latter expression was based on
a well defined classical trajectory parametrized by $\tau$
whereas in the present case, it is momentum conservation that has 
been exactly taken into account.

However, it is precisely this conservation law 
that re-introduces the notion of classical trajectory (exactly like in quantum gravity
the Wheeler-DeWitt constraint re-introduces the notion of 
time\cite{wdw}).
To explicitize this, 
we develop the phase and the norm of
the {\it integrand} of eq. (\ref{tildB2}) in powers of $\om$ and $\Delta m$.
To first order in $\Delta m$ and $\om$, the phase $\varphi(t)$
 is {\it equal}, up to a constant, to the phase of the 
integrand of eq. (\ref{B1g}). Indeed, one has
\begin{eqnarray}
\varphi(t) &=& \int^t dt^\prime \left[ p(M,k - \om, t^\prime) - p(m, k, t^\prime)
\right] + \om t 
 \nonumber\\ 
&\simeq&\Delta m \  \partial_M  \int^t dt^\prime  p(M,k , t^\prime) - 
\om \ \partial_k  \int^t
dt^\prime p(M,k , t^\prime) + \om t  \nonumber\\ &\simeq& 
\Delta m \Delta \tau(t) - \om ( \Delta z_k(t) - t)
\label{phasa}
\end{eqnarray}
$\Delta \tau(t)$ and $\Delta z_k(t)$ are  
the lapses of proper time and of space
evaluated along a uniformly accelerated trajectory characterized by $k$.
These relations may be checked explicitly by computing the integrals 
using eq. (\ref{KGE}).

However, for establishing their universal validity, it is 
appropriate to realize that those relations are nothing but 
Hamilton-Jacobi relations. Indeed, these are
\ba
\partial_M S_{cl.}(M, k, t ) &=& \partial_M\! \int^t \! dt^\prime  p(M,k, t^\prime) =  \Delta \tau(t)
\label{nn}
\\
 \partial_k S_{cl.}(M, k, t ) &=&  \partial_k\! \int^t \! dt^\prime  p(M,k , t^\prime) = \Delta z_k(t)
\label{nnn}
\ea
Thus, whatever is the nature of the external field which brings
the system into constant acceleration, the first two terms of eq. (\ref{phasa})
will always be found. More importantly and more generally, 
whatever is the problem one is considering,
upon using WKB wave functions for describing ``heavy'' degrees of freedom,
the transition amplitudes among ``light'' degrees will agree with the BFA expressions
upon developing the amplitudes to first order in the light changes.   
Indeed, in that approximation,
one must also neglect the dependence in $\om$ and $\Delta m$ 
in the denominator of eq. (\ref{tildB2}). Then the measure is
$dt/p(M,k, t) = d\tau /M$. Thus,
one has, as announced,
\begin{eqnarray}
\tilde B(M, m, k, \om) &=& {-i \tilde g M}  \int^{+\infty}_{-\infty} d\tau 
e^{i\Delta m \tau} {e^{-i \om e^{-a\tau}/a} \over \sqrt{4 \pi \om}}
\times \left( e^{-i \om k/E } \right)
\nonumber\\
 &=& \left[ {\tilde g M \over ga} \right]
B(\Delta m, \om, a)\ \times \left( e^{-i  \om k/E } \right)
\label{tildB3}
\end{eqnarray}

Very important is the fact that the momentum $k$ introduces only
a phase shift 
with respect to the BFA amplitude $B(\Delta m, \om, a)$.
Thus any normalized superposition of modes $\psi_{m,k}$ specifying the initial
wave function of the ``heavy'' system will give rise to the same
probability to emit of photon of energy $\om$.  Moreover the ratio of
the square of $\tilde B(M, m, k, \om)$ and
$\tilde A (M, m, k, \om)$ will necessarily 
satisfy eq. (\ref{ratio2}). Therefore, under these approximations,
the two level ion thermalizes exactly as in the no-recoil case.
Moreover, the total energy emitted by the ``recoiling'' 
ion also equals the corresponding energy evaluated at the BFA.
The reason is simply that the total energy is a function of the norm of $B(M, m, \om, k)$
only\cite{rec}.

The lesson of this comparison of amplitudes is the following.
When both the WKB approximation for the ``heavy'' system 
and a first order expansion in the light changes are valid,
Hamilton-Jacobi equations guarantee that the norm of the transition
amplitudes agree. Therefore all physical quantities that are functions of these
norms only will automatically agree as well.

However, there are both conceptual and numerical differences between the
amplitudes computed in the two frameworks.
For instance the phase of the amplitudes do not agree even to first order in
$\om$ and $\Delta m$. 
We shall illustrate these differences by first considering the 
stationary phase condition in both cases and then by analyzing the 
consequences of the phase shifts in the determination of the flux
emitted by the atom. 

In this present framework, the stationary point $t^*$ 
of the integrand of eq. (\ref{tildB2}) is at
\begin{equation}
p(M,k - \om, t^*) - p(m, k, t^*) + \om = 0
\label{sp2}
\end{equation}
This is the conservation law of the Minkowski energy. 
Instead, in the Unruh framework, from the first line of eq. (\ref{B1g}), one finds
\be
\Delta m + \om e^{-a \tau} = 0
\label{saddtau}
\ee
which is the resonance condition in the accelerated frame, i.e.
conservation of Rindler energy.
This difference also arise in gravitational situations, see \cite{wdw}:
 Upon dealing with gravity described at the 
BFA, one finds that matter processes satisfy energy conservation in the
given background; this is the equivalent of eq. (\ref{saddtau}). 
And upon abandoning the BFA and solving twice Einstein's equations, 
one verifies that the resonance
condition involves the energy of gravity, exactly like 
eq. (\ref{sp2}) contains the difference of two heavy energy $p(M, k, t)$.

Of course, 
the two versions of energy 
conservation must coincide in the limit $\Delta m/ M \to 0$. 
Indeed, by taking the
square of eq. (\ref{sp2}) and using eq. (\ref{KGE}) one gets
\begin{equation}
{M^2 - m^2 \over 2 } = \om \left[ ( k - \om +Et^*) - p(M,k - \om, t^*)\right]
\label{sp2b}
\end{equation}
Introducing once more the proper time of the heavy ion $M$,  
eq. (\ref{eqmot}), one finds
\begin{equation}
\Delta m (1 - \Delta m / 2M) = - \om e^{-a \tau^*}\quad\quad {\mbox{QED}}
\label{sp2c}
\end{equation}

The second point we wish to make is the following.
There is a strict relation between conservation 
of momentum, eq. (\ref{calB}), and energy, eq. (\ref{sp2}),
and the modifications of the properties of the emitted fluxes, see \cite{rec}
for more details. Indeed, the two level ion constantly loses energy and momentum
in accordance with these conservation laws.
Then the trajectory of the ion
drifts from orbits to neighboring ones in $t$ and $z$ corresponding
to laters times and greater $z$. The total change in the time of 
the turning point is $ E \Delta t_{t.p.}= \sum_i \om_i$, 
i.e. it is proportional to the total momentum lost. 
One also verifies that the total change in position is $\sum_i \om_i/E$.
These successive changes of hyperbolae lead to the 
decoherence of the emissions causing these changes since
the very peculiar phases obtained in the BFA framework are washed out by the recoils. 
Then, both ``unphysical'' properties obtained in that framework, namely negativity
of the energy density during arbitrary large proper times\cite{grove}-\cite{MPB} 
and singular behavior
on the horizon\cite{Unruh92}\cite{mapa1}, are eliminated.

There is a third point that can be addressed before considering 
corrections to the WKB approximation. It concerns the possibility
in computing corrections of the equilibrium ratio, eq. (\ref{ratio3}), 
induced by higher order terms in $\Delta m / M$. 
This problem shall be discussed 
elsewhere.


\section{The amplitudes in second quantization}

In the former Section, we used WKB wave functions which are valid
approximate solutions of eq. (\ref{KGE2}) when $M^2/E \gg 1$.
Then, positive and negative energy solutions completely
decouple. This is no longer true when one deals with the
exact solutions of eq. (\ref{KGE2}). 
In this new case indeed, one must introduce two 
sets of modes ($\chi^{in}_M, \chi^{out}_M$) which only asymptotically, i.e. for $t \to \pm \infty$,
define particle states.
That the two states do not coincide (they are related by a 
linear (Bogoliubov) transformation)
 indicates that the initial vacuum $\ket{0, M, in}$ associated with the initial set $\chi^{in}_M$
 spontaneously decays and that particles will be find at $t \to \infty$.  
 In the present case of a constant electric field, 
one finds that the mean
number of quanta of momentum $k$ is given by 
\be
N_M = \vert \beta_M \vert ^2 = e^{- \pi M^2/E}
\label{sc}
\ee
where $\beta_M$ is the Bogoliubov coefficient, i.e. the amount of negative energy 
{\it out} solution in a purely positive {\it in} solution, see {\it e.g.} \cite{GO}.

Using Feynman rules, one can re-calculate the
ratio of the transition rates
to emit a photon starting from the ground state ($m$) and from the excited state ($M$)
by taken into account vacuum instability
with respect to pair creation of both ion fields, see Nikishov\cite{Niki}.
This ratio
turns out to be intimately related\cite{BPS}\cite{suh} 
to eq. (\ref{sc}) in that it is equal to
\be
\vert{ {\cal B}(M, m, p, \om) \over {\cal A}(M, m, p, \om)} \vert^2= 
e^{-\pi (M^2-m^2) /E} = { N_M \over N_m }
\label{exactT}
\ee

To re-calculate this
ratio, we analyze
the amplitude $ {\cal B}(M, m, k, \om)$
to emit a massless quantum of energy $\om$ starting from the state $m$.
This amplitude strictly
corresponds to the amplitude $\tilde B(M, m, k, \om)$ of eq. (\ref{calB}).
To first order in $\tilde g$, 
it is given by
\begin{eqnarray}
&&\delta  (k-k'-{\om} ){\cal B}(M, m, k, \om) \nonumber \\
&&\quad \quad \quad = \bra{1_\om} 
\bra{1_{k'}, M, out} \bra{0, m, out} 
e^{-i \int dt H_{\psi \phi}} \ket{1_{k}, m, in}\ket{0, M, in}\ket{0}
\nonumber \\
&&\quad \quad  \quad =- \delta(k-k'- {\om})\ i \tilde g M^2
\left( Z_M Z_m 
\alpha_M^{-1} \alpha_m^{-1}\right)
\nonumber \\
&&\quad \quad \quad \quad \quad \times 
\int^\infty_{-\infty} dt \; \chi^{{in} *}_{M, k- \om}(t) \; \chi^{out}_{m, k}(t) \; 
{e^{i\om t } \over \sqrt{4 \pi \om} }
\label{ov}
\end{eqnarray}
The factor $Z_M Z_m $ is the product of the overlaps of the {\it in}
and {\it out} vacuum states of both charged fields.
$\alpha_{M(m)}$ is a coefficient whose norm is given by $\sqrt{ 1 +\vert 
\beta_{M(m)}\vert^2}$.
These factors take into account the fact that the
scattering process happens in the presence of pair production of charged quanta.
They all reduce to one in the WKB limit, $M^2/E \to \infty$.

This amplitude can be exactly evaluated in terms of the 
integral representations of the $\chi$ modes, see \cite{suh}.
One obtains,
\ba
&&{\cal B}(M, m, k, \om)
=-i {\tilde g  M^2 \over 2E}
\ \Gamma({-i(M^2-m^2)/ 2E})
\nonumber\\
 && \quad\quad\quad\quad \times  
\ {(\om)^{{i(M^2-m^2)/2E}}\over \sqrt{4 \pi \om}}
\ e^{-\pi (M^2-m^2)/2 E}
\nonumber\\
&& \quad\quad\quad\quad \times  e^{i(\om p - \om^2 /2)/ E}\
\left[ \Gamma(i{M^2 \over 2E} +{1\over 2})  {e^{\pi M^2 /2 E}   
(E/2)^{-iM^2/2E} \over \sqrt{2 \pi}} \right]\quad\quad\quad
\label{App51}
\ea
This expression should be compared with eq. (\ref{B1g}).
As in that case, the determination 
of the equilibrium population requires only to know the ratio of the 
transition rates. And as in that case, this ratio can be
obtained effortless from the analytical behavior of the
amplitude in $\om$, since
 ${\cal A}^*(M, m, k, \om) = {\cal B}(M, m, k, e^{-i \pi}  \om)$.
(Notice that this analytical behavior 
in $\om$ encodes the stability condition of the Minkowski
vacuum of the radiation field. Notice also that Hawking radiation can 
be derived in similar terms\cite{Unr}.) From this relation 
and eq. (\ref{App51}), one   immediately deduces that
the ratio of the transition rates
satisfy eq. (\ref{exactT}).

Therefore we have proven that
the equilibrium probabilities defined by radiative 
processes are equal to those
defined by the Schwinger process.
However, the direct proof that both processes are intimately 
related follows from the fact that 
their respective amplitudes are interchanged by crossing symmetry,
see \cite{suh}. Indeed what corresponds to a pair creation diagram
in the direct channel, describes a radiative transition in the ``crossed''
channel. 

To conclude this Section, we analyze the relationships
between this formulation of radiative processes with the former
ones studied in Section 2 and 3. 
%
First notice that in the limit $M^2/E \to \infty$ at fixed $M-m= \Delta m$ and $M/E=1/a$, 
the {\it integrand} of eq. (\ref{ov}) tends uniformly 
to the {\it integrand} based on WKB expressions in eq. (\ref{tildB2}).
This indicates once more that the rates agree because matrix elements
defining transition amplitudes 
can be put in strict correspondence, at fixed quantum number.

Secondly, it should not have escaped the reader that eq. (\ref{exactT}) differs
from the Boltzmanian ratio found using the BFA, i.e. eq. (\ref{ratio2}).
Indeed, in order to make contact with this thermal 
{\it canonical} distribution, one must consider the limit $\Delta m <\!\!< M$.
Then, in the present second quantized framework, 
the concepts of acceleration and temperature are brought to bear
for the first time.
They both appear through a first order change in $\Delta m$.
This emergence of classical concepts
bears many similarities with statistical
mechanics since it is also through a first order change in the energy 
that the concept of temperature arises from {\it microcanonical}
ensembles. 

To bring about this contact with statistical
mechanics, it is most instructive
to use again Hamilton-Jacobi equations but applied this time to Euclidean 
dynamics.
Indeed relationships with both eq. (\ref{nn})
and black hole thermodynamics will become clear.
We remind the reader that the Schwinger pair creation amplitudes may be
understood from the Euclideanized version of the problem. 
In fact, it is $S_{euclid} =  \pi M^2 / E$, 
the classical action 
to complete a Euclidean (closed) orbit, which determines their rate, 
see eq. (\ref{sc}). From this action, one computes the Euclidean proper time 
necessary to complete this orbit. It is given by
 \be
\tau_{euclid} = \partial_M S_{euclid} = \partial_M 
\left( { \pi M^2 \over E} \right)  = \left( { a \over 2 \pi}\right) ^{-1}
\label{prop}
\ee
It equals the inverse Unruh temperature. 
Thus, to first order in $\Delta m$, 
it is meaningful to write eq. (\ref{exactT}) as 
\be
\vert{ {\cal B}(M, m, k, \om) \over {\cal A}(M, m, k, \om)} \vert^2
\simeq  e^{- {\Delta m} \partial_M S_{euclid}} =  e^{- {\Delta m} 2 \pi /a}= 
\vert{ {B}(\Delta m, a, \om) \over { A}(\Delta m, a, \om)} \vert^2
\label{equal}
\ee
This shows that the Unruh process can be viewed as an
infinitesimal ratio of two Schwinger processes.
Moreover the instanton action $S_{euclid}$ acts as an {\it entropy}
in delivering the Unruh temperature. Indeed, 
\be
 \tau_{euclid} \Delta m = \Delta S_{euclid}
\label{firstl}
\ee
is the first law of thermodynamics\footnote{
For the skeptical reader, we add that this analysis has been enlarged\cite{apple}
 by replacing the massless neutral field by a charged field. Then the new equilibrium 
ratio is determined by an extended thermodynamical relation in which the work 
done by the electric field contributes as well. The analogy
with the thermodynamics of charged black holes is manifest.}.

In view of the similarities between Unruh effect and black hole
radiation, it is inviting to inquire
about the relationship between this ``instantonic'' entropy and the
gravitational entropy of black holes whose variation 
determines Hawking temperature.
This is the subject of next Section.

\section{The amplitudes in the presence of gravity}

This short Section is of a heuristic character and the level of 
mathematical rigor  
will be lowered simply because Quantum Gravity does
not exist. Indeed, to describe quantum transitions in which gravity does play
an active role, i.e. in which the BFA for gravity has been abandoned,
new approximations should be adopted. The type of approximations we shall
need and use are quite similar to those we
used in Section 3 for describing the system's trajectory:
Gravitational quantum effects will approximatively be taken into account
like momentum recoils were taken into account in eq. (\ref{tildB2}), i.e. 
in matrix elements describing transition amplitudes of light (matter) 
degrees of freedom, the initial and final wave functions of the 
heavy system (gravity) bring their own classical action, see \cite{wdw}.

To illustrate these aspects, we start the discussion by recalling the results 
of \cite{(11)}-\cite{(4)}.
In these references, the probability of creation charged black holes
in a constant electric field was estimated by making use of two 
hypothesis. First the authors required
that the Euclidean manifold describing the instanton
responsible for the creation be regular.
Secondly they assumed that the probability of creation depends on the 
instanton action computed from the Hilbert-Einstein like 
 $S_{euclid}$,
the classical action 
to complete a Euclidean orbit, determined the production rate in
eq. (\ref{sc}).
Using these hypothesis, they found that the probability to 
produce a pair of black holes characterized by the mass
 $M$ and charge $Q$ is 
\begin{equation}
P_{M,Q} = e^{ \delta {\cal A}_{acc}/ 4G}  \times e^{ {\cal A}_{BH}/ 4G}
\label{27e}
\end{equation}
 
${\cal A}_{BH} (M,Q)$
is the area of the black hole horizon.
In this 
expression, $e^{ {\cal A}_{BH}/ 4G}$ furnishes the density of black hole 
states with mass $M$ and charge $Q$ thereby confirming the 
Bekenstein interpretation
of ${\cal A}_{BH}/4G= S_{BH}$ as being the black hole entropy.

$\delta {\cal A}_{acc} (M, Q, E) $ is the {\it change} of the area of the 
acceleration horizon induced by the creation of the black hole 
pair\cite{HH}.

The domain of the one parameter 
family (i.e. the values of $M$ and $Q$ which satisfy the regularity condition)
which can be compared with the Schwinger mechanism
is the one in which the black holes radius is much smaller than
 the inverse acceleration, i.e. in the point particle limit. 
This limit corresponds to the limit $G \to 0$. In that case, one finds
\be
{  \delta {\cal A}_{acc}  \over 4G} =
- \pi M^2 / Q E = - S_{euclid}
\label{E1}
\ee
This is the usual instanton action to create charged particles.

First notice that it is independent of $G$, as it should be. 
Indeed, the limit $G \to 0$ in the present gravitational case strictly
corresponds to the limit $\Delta m/M \to 0$ in Sections 3 and 4. 
In that case, we saw that the transition probabilities coincide with the BFA
probabilities since the integrands of the amplitudes 
leading to these probabilities agree up to a phase.
Similarly, by turning on gravity and then taking the limit $G \to 0$,
one must recover the BFA probabilities. That this is the case, justify
a posteriori the assumptions used in \cite{(4)}.

Secondly, eq. (\ref{E1}) and
eq. (\ref{firstl}) show 
that $S_{euclid}$ acts as a gravitational entropy. Indeed, eq. (\ref{firstl}) 
can be rewritten as
\be
{2 \pi \over \kappa }\Delta m = {\Delta {\cal A}_{acc} \over 4G}
\label{firstl2}
\ee
where $\kappa = a$ is the surface gravity of the accelerating horizon
measured along the trajectory $z^2 - t^2 = 1/a^2$.
$\Delta {\cal A}_{acc}$ is the change of the accelerating horizon area 
associated with the replacement of the heavy mass $M$ by the lighter one
$m$\cite{gravitinst}.
This rewriting strongly suggests that changes in area divided by $4G$
determine equilibrium distributions of accelerated systems
according to the first law of horizon thermodynamics.

More details on these aspects can be found in \cite{gravitinst}\cite{apple}.

\section*{Acknowledgments}

I am grateful to Serge Massar and Ted Jacobson for the
helpful discussions we had during the ``MG8'' and ``The Internal 
Structure of Black Holes and Space-time Singularities'' meetings. I 
wish also to thank the organizers of these meetings for 
inviting me.


\begin{thebibliography}{999}

\bibitem{israel} W. G. Anderson, P. R.
       Brady, Werner Israel, and S. M. Morsink, Phys. Rev. Lett. 70 (1993) 1041

\bibitem{ori} 
L. M. Burko and  A. Ori, Phys. Rev. Lett. 74 (1995) 1064

\bibitem{t} S. P. Trivedi, ``Semiclassical extremal black holes'', hep-th/9211011

\bibitem{teddy} T. Jacobson, ``Semiclassical decay of near extremal black holes'', 
hep-th/9705017

\bibitem{Hawk} S. W. Hawking, Commun. Math. Phys. {\bf  43} (1975) 199

\bibitem{Bard} J. M. Bardeen, Phys. Rev. Lett. {\bf 46} (1981) 382

\bibitem{PP} R. Parentani and T. Piran, Phys. Rev. Lett. {\bf 73} (1994) 2805

\bibitem{Massar2}  S. Massar, Phys. Rev. D {\bf  52} (1995) 5861

\bibitem{THooft} G. 't Hooft, Nucl. Phys. B {\bf 256} (1985) 727

\bibitem{Jacobson} T. Jacobson, Phys. Rev. D {\bf 44} (1991) 1731 and
 {\bf 48} (1993) 728

\bibitem{unruh94} W. G. Unruh, Phys. Rev. D {\bf  51} (1995) 2827

\bibitem{bmps} R. Brout, S. Massar, R. Parentani and Ph. Spindel,
Phys. Rev. D {\bf  52} (1995) 4559

\bibitem{mapa2} S. Massar and R. Parentani, Phys. Rev. D {\bf  54} (1996) 7431

\bibitem{Verl3} Y. Kiem, H. Verlinde, E. Verlinde {\it Quantum Horizons and
Complementarity.}
CERN-TH-7469-94, hep-th/9502074

\bibitem{Unr}  W. G. Unruh, Phys. Rev. D {\bf  14} (1976) 870
 
\bibitem{rec} R. Parentani, Nucl. Phys. B {\bf 454} (1995) 227

\bibitem{BPS} R. Brout, R. Parentani and
Ph. Spindel, Nucl. Phys. B {\bf 353 } (1991) 209

\bibitem{wdw} R. Parentani, Nucl. Phys. B. {\bf 492} (1997) 475, 501,
gr-qc/9610044 and 45.

\bibitem{Unruh92} W. G. Unruh, Phys. Rev. D {\bf 46} (1992) 3271

\bibitem{mapa1} S. Massar and R. Parentani, Phys. Rev. D {\bf  54} (1996) 7426

\bibitem{schw} J. Schwinger, Phys. Rev. {\bf 82} (1951) 664

\bibitem{suh} S. Massar and R. Parentani, Phys. Rev. D {\bf  55} (1997)
3603

\bibitem{Niki} A. I. Nikishov, Sov. Phys. JETP {\bf 30} (1970) 660


\bibitem{apple} C. Gabriel, Ph. Spindel,  S. Massar, R. Parentani,
``Interacting Charged Particles in an Electric Field and the Unruh Effect'',
hep-th/9706030

\bibitem{(11)} G. W. Gibbons, in ``Fields and Geometry'', Proceedings
  of the 22nd Karpacz Winter School of Theoretical Physics, Karpacz,
  Poland, 1986, edited by A. Jadczyk (World Scientific, Singapore, 1986)

\bibitem{(1)} D. Garfinkle and A. Strominger, Phys. Lett. B {\bf 256} 
(1991) 146

\bibitem{(2)} S. W. Hawking and S. Ross, Phys. Rev. D {\bf 52} (1995) 5865

\bibitem{(4')}
F. Dowker, J. P. Gauntlett, S. B. Giddings and G. T. 
Horowitz, Phys. Rev. D {\bf 50} (1994) 2662

\bibitem{(4)} S. W. Hawking, G.T. Horowitz and S. Ross, Phys. Rev. D 
{\bf 51} (1995) 4302

\bibitem{gravitinst} S. Massar and R. Parentani, Phys. Rev. Lett. {\bf 78} (1997) 3810 

\bibitem{jac} T. Jacobson, Phys. Rev. Lett. {\bf 75} (1995) 1260

\bibitem{Full} S. A. Fulling, Phys. Rev. D {\bf  7} (1973) 2850


\bibitem{UnrW} W. G. Unruh and R. M. Wald, Phys. Rev. D {\bf 29} (1984) 1047

\bibitem{grove} P. Grove, Class. Quant. Grav. {\bf  3} (1986) 801

\bibitem{RSG} D. Raine, D. Sciama and P. Grove, Proc. R. Soc. A {\bf 435} (1991)
205

\bibitem{MPB} S. Massar, R. Parentani and R. Brout,
Class. Quant. Grav. {\bf 10 } (1993) 385

\bibitem{BL} J. S. Bell and J. M. Leinaas, Nucl. Phys. B {\bf 212} (1983) 131


\bibitem{BMPPS} R. Brout, S. Massar, R. Parentani, S. Popescu and
Ph. Spindel,  Phys. Rev. D {\bf 52} (1995) 1119

\bibitem{GO} R. Brout, S. Massar, R. Parentani and Ph. Spindel,
Phys. Rep.  {\bf 260}  (1995) 329

\bibitem{HH} S. W. Hawking and G.T. Horowitz, Class. Q. Grav. {\bf
    13} (1996) 1487 



\end{thebibliography}
\end{document}